\documentclass[amsmath,12pt]{article}

\begin{document}

\begin{titlepage}

\begin{center}

{\Large \bf
Relativistic action-at-a-distance interactions: Lagrangian and Hamiltonian
to terms of second order}
\vskip .6in

Domingo J. Louis-Martinez
\vskip .2in

Department of Physics and Astronomy,\\ University of British
Columbia\\Vancouver, Canada, V6T 1Z1 

martinez@physics.ubc.ca

\end{center}
 
\vskip 3cm
 
\begin{abstract}
Relativistic systems of particles interacting pairwise at a distance
(interactions not mediated by fields) in flat spacetime are studied. 
It is assumed that the interactions propagate at
the speed of light in vacuum
and that all masses are scalars under Poincar\'e transformations.
The action functional of the theory depends on multiple times (the proper times
of the particles).
In the static limit, the theory has three components: a linearly rising
potential, a Coulomb-like
interaction and a dynamical component to the Poincar\'e invariant mass.
In this Letter we obtain explicitly,
to terms of second order, the Lagrangian and the Hamiltonian
with all the dynamical variables depending on a single time.
Approximate solutions of the relativistic two-body problem are
presented. 
\end{abstract}

PCAS No. 03.30+p

\end{titlepage}

After the discovery of the action-at-a-distance formulation of
electrodynamics\cite{wheeler} - \cite{barut}, several relativistic non-instantaneous 
action-at-a-distance
theories have been investigated \cite{dettman} - \cite{friedman}. Instantaneous
action-at-a-distance formulations have been studied using a variety
of approaches \cite{hill} - \cite{droz}.

In a recent paper \cite{domingo2}
the relativistic action-at-a-distance approach of Wheeler and Feynman
\cite{wheeler} - \cite{barut}
was extended in order to explore what other types of interparticle
interactions
are allowed in special relativity. It was assumed that the interactions propagate
at the speed of light and that the theory can be described by an action
that does not depend on the
four-vector accelerations or on higher derivatives.
A general action functional depending only
on the four-vector velocities and relative positions of the particles
was investigated. 
Assuming that the Poincar\'e invariant parameters that
label successive events along the world lines of the particles can be
identified with the particles' proper times and that all masses are scalars,
the most general form of the interaction
terms in the action was determined explicitly \cite{domingo2}.

The theory presented in \cite{domingo2} is described by an action
which depends
on multiple times.
Quantum mechanics, on the other hand, is based on the
hamiltonian formalism (or the Dirac hamiltonian formalism for constrained
systems\cite{dirac} - \cite{henneaux}), which is built on the assumption that
one can choose a single
time variable to describe the evolution of the whole system. The lagrangian
(or covariant) quantization procedure \cite{bv} - \cite{bvdom}
also assumes a single time variable.

One may attempt to transform a theory
with multiple time variables into an equivalent theory with just one time
variable, in a given inertial reference frame,
using Taylor series expansions involving the particles' present
motions \cite{kerner}. The problem of finding a single time variable
to describe a relativistic system of interacting particles has been investigated by
several authors \cite{horwitz2} - \cite{uri}.

In this Letter we present a reformulation of the 
theory given in \cite{domingo2} as an action-at-distance theory with
a single time variable. 
We use series expansions up to terms of second order
($\frac{v^{2}}{c^{2}}$).

We obtain the approximate Lagrangian for this theory.
Approximate expressions for the total energy, total linear momentum
and total angular
momentum are presented. We also obtain the Hamiltonian to terms of second
order.

We present a formula for the Hamiltonian of two
interacting particles in the center of momentum reference frame.
Using this result, direct predictions for the energy spectrum can be made.

We also present approximate circular orbits solutions
of the relativistic two-body problem.

It is well known that an isolated system of $N$ particles interacting
electromagnetically may be described, to terms of second order, 
by the Darwin Lagrangian \cite{darwin},\cite{landau}:

\begin{eqnarray}
L_{Darwin} & = & - c^{2}\sum\limits^{}_{i} m_{i}
+ \frac{1}{2} \sum\limits^{}_{i} m_{i} v_{i}^{2}
+ \frac{1}{8 c^{2}}\sum\limits^{}_{i} m_{i} v_{i}^{4}
- \frac{1}{2} \sum\limits^{}_{i} \sum\limits^{}_{j \neq i}
\frac{e_{i} e_{j}}{r_{ij}}
\nonumber\\
& & + \frac{1}{4 c^{2}}
\sum\limits^{}_{i} \sum\limits^{}_{j \neq i}
\frac{e_{i} e_{j}}{r_{ij}}
\left((\vec{v}_{i} \vec{v}_{j}) 
+(\vec{n}_{ij} \vec{v}_{i})(\vec{n}_{ij} \vec{v}_{j})\right)
\label{P1}
\end{eqnarray}

\noindent where, $m_{i}$ and $e_{i}$ are the mass and electric charge of
particle $i$ ($i = 1,2,...,N$), $\vec{v}_{i}$ its velocity,
$\vec{r}_{ij} = \vec{r}_{i} - \vec{r}_{j}$ the relative position of
particle $i$ with respect to particle $j$ and $c$ the speed of light.
In (\ref{P1}), $\vec{n}_{ij} \equiv \frac{\vec{r}_{ij}}{r_{ij}}$.

A generalization of the Darwin Lagrangian that allows the inclusion of other types
of interactions was proposed by Woodcock and Havas in the 70's \cite{havas}.
Other formulations have recently been studied \cite{lusanna}.
For quantum chromodynamics, quark-antiquark interactions to terms of second
order have been studied by several authors \cite{miller} -\cite{sazdjian}.
Gravitational interactions, to terms of second order, are described by
the Einstein-Infeld-Hoffmann Lagrangian \cite{einstein}. 

To terms of fourth order ($\frac{v^4}{c^4}$),
ignoring the radiation
effects (which for Faraday-Maxwell electrodynamics appear at third order)
the Lagrangian for an isolated
system of N particles interacting electromagnetically
was obtained in \cite{smoro}, \cite{ohta}.  For General Relativity,
gravitational radiation effects occur at fifth order. To terms of fourth
order, the Lagrangian for an isolated system of N particles with gravitational
interactions was derived in \cite{damour}.
In recent years, for the gravitational two-body problem, results to sixth
order (3rd post-Newtonian approximation) and higher have been obtained
\cite{blanchet}.

The Lagrangian (\ref{P1}) depends only on the velocities and the relative
positions of the particles, which are functions of a single time variable 
in a given inertial reference frame $K$. The Darwin Lagrangian can be derived either from
Faraday-Maxwell's field theory of electrodynamics \cite{landau} or from
the relativistic action-at-a-distance theory of Wheeler and Feynman
\cite{anderson}.

Wheeler-Feynman's fully relativistic action-at-a-distance theory of
electrodynamics is described by the Fokker action \cite{wheeler} -
\cite{barut}:

\begin{equation}
S = - \sum\limits^{}_{i} m_{i} c \int d\lambda_{i} \zeta_{i}^{\frac{1}{2}} 
- \frac{1}{2 c}\sum\limits^{}_{i}  \sum\limits^{}_{j\neq i}
e_{i} e_{j}
\int \int d\lambda_{i} d\lambda_{j}
\xi_{ij}
\delta\left(\rho_{ij}\right)
\label{P2}
\end{equation}

In (\ref{P2}), $\lambda_{i}$ is a Poincar\'e invariant parameter that labels
the events along the world line
$z^{\mu}_{i}(\lambda_{i})$
of particle $i$ in flat Minkowski spacetime and,

\begin{equation}
\zeta_{i} = \dot{z}_{i}^{2}
\label{P3}
\end{equation}

\begin{equation}
\xi_{i j}= \left(\dot{z}_{i} \dot{z}_{j}\right) 
\label{P4}
\end{equation}

\begin{equation}
\rho_{i j}= \left( z_{i} - z_{j}\right)^{2} 
\label{P5}
\end{equation}

\noindent are scalars under Poincar\'e transformations.

The metric tensor: $\eta_{\mu\nu}= diag(+1,-1,-1,-1)$.
In (\ref{P3}, \ref{P4}),
$\dot{z}^{\mu}_{i} = \frac{d z^{\mu}_{i}}{d \lambda_{i}}$.

The Dirac delta function in (\ref{P2}) accounts for the interactions
propagating at the speed of light forward and backward in time.

In action-at-a-distance electrodynamics the interactions carry energy and
momentum to and from the particles and may simulate a field between them.
However, in the absence of electrically charged particles outside the system,
this fictitious field cannot carry energy or momentum into or away from the
system. If there are no other electrically charged particles in the universe,
the total energy, linear momentum and angular momentum
of a system of point particles are
conserved (assuming that only electromagnetic interactions take place).

In \cite{domingo2},
a general action functional depending
on the four-vector velocities and relative positions of the particles
was investigated. Assuming that the Poincar\'e invariant parameters that
label successive events along the world lines of the particles can be
identified with the particles' proper times in flat spacetime and that all 
masses are scalars,
the most general form of the interaction
terms in the action was determined explicitly.

The theory obtained in \cite{domingo2} is described by the following
action functional:

\begin{equation}
S = - \sum\limits^{}_{i} m_{i}c\int d\lambda_{i} \zeta_{i}^{\frac{1}{2}}  
- \frac{1}{2 c} \sum\limits^{}_{i}  \sum\limits^{}_{j\neq i} g_{i} g_{j}
\int \int d\lambda_{i} d\lambda_{j}  \delta\left(\rho_{ij}\right)
\left( \alpha \gamma_{ij} \gamma_{ji} + \beta \xi_{ij}
+ \gamma \zeta_{i}^{\frac{1}{2}} \zeta_{j}^{\frac{1}{2}}\right)
\label{P6}
\end{equation}

In (\ref{P6}), $\alpha$, $\beta$ and $\gamma$ are constants, and

\begin{equation}
\gamma_{i j}= \left(\dot{z}_{i} (z_{j} - z_{i})\right) 
\label{P7}
\end{equation}

The Poincar\'e invariant parameters $\lambda_{i}$ are identified with
the proper times ($\tau_{i}$)
of the particles ($\lambda_{i} = s_{i} = c \tau_{i}$)
\cite{domingo2}:

\begin{equation}
d\lambda_{i}^{2} = \eta_{\mu\nu} d z^{\mu}_{i} d z^{\nu}_{i}
\label{P7a}
\end{equation}

In the static limit the theory has three components: a linearly rising
potential, a Coulomb-like interaction and a dynamical component to the
Poincar\'e invariant mass.

There is
strong experimental evidence indicating that, for large separations, the interactions
between quarks can be effectively described by a linearly rising potential
\cite{lucha}, \cite{olsson}.  Several relativistic generalizations of a linearly rising
potential have been studied \cite{katz} - \cite{weiss}.
From quantum chromodynamics, 
it has been shown that the quark-antiquark bound states are
effectively described
by a static potential, which is a sum of a
linearly rising potential and a Coulomb-like interaction
\cite{allen}. It would be interesting to study the effect of a variable
scalar mass for quarks, as given by the formula (\ref{P9}) below
or in some other form.

From the action (\ref{P6}), using the variational principle,
one can obtain the relativistic equations of
motion for a system of $N$ interacting particles \cite{domingo2}:

\begin{equation}
\bar{m}_{i} \ddot{z}^{\mu}_{i} = \bar{K}^{\mu}_{i}
\label{P8}
\end{equation}

In (\ref{P8}) $\bar{m}_{i}$ is the dynamical mass
of particle $i$ ($i = 1,2,...,N$):

\begin{equation}
\bar{m}_{i} = m_{i} + \gamma \frac{g_{i}}{c^2} \sum\limits^{}_{j\neq i} g_{j}
\int d s_{j} \delta\left(\rho_{ij}\right)
\label{P9}
\end{equation}

\noindent and $\bar{K}^{\mu}_{i}$ is the four-vector force acting on
particle $i$  ($i=1,2,...,N$):

\begin{equation}
\bar{K}^{\mu}_{i} = g_{i} \left(F^{\mu\nu}_{i} \dot{z}_{i\nu}
+ \Gamma^{\mu}_{i \alpha\beta} \dot{z}^{\alpha}_{i}\dot{z}^{\beta}_{i}\right)
\label{P10}
\end{equation}

\noindent where $F^{\mu\nu}_{i}$ is an antisymmetric tensor
($F^{\nu\mu}_{i} = - F^{\mu\nu}_{i}$) 
and $\Gamma^{\mu}_{i\alpha\beta}$ is a symmetric tensor
($\Gamma^{\mu}_{i\alpha\beta} = \Gamma^{\mu}_{i\beta\alpha}$) in flat
spacetime. They are given by the expressions \cite{domingo2}:

\begin{eqnarray}
F^{\mu\nu}_{i} & = & \sum\limits^{}_{j\neq i} \frac{g_{j}}{c^{2}} \int d s_{j}
\frac{\delta\left(\rho_{ij}\right)}
{\gamma_{ji}^{2}} \nonumber\\
& & \left[ \alpha \gamma_{ji}^{2}
\left[(z^{\mu}_{i} - z^{\mu}_{j})\dot{z}^{\nu}_{j} -
\dot{z}^{\mu}_{j} (z^{\nu}_{i} - z^{\nu}_{j})\right]  \right. \nonumber\\
& & + \beta
\left[\left((z^{\mu}_{i} - z^{\mu}_{j})\dot{z}^{\nu}_{j} -
\dot{z}^{\mu}_{j} (z^{\nu}_{i} - z^{\nu}_{j})\right)
\left(1 - (\ddot{z}_{j} (z_{i} - z_{j}))\right) \right. \nonumber\\
& & + \left. \left. \left((z^{\mu}_{i} - z^{\mu}_{j})\ddot{z}^{\nu}_{j} -
\ddot{z}^{\mu}_{j} (z^{\nu}_{i} - z^{\nu}_{j})\right)
\gamma_{ji}\right]\right]
\label{P11}
\end{eqnarray}

\begin{eqnarray}
\Gamma^{\mu}_{i\alpha\beta} & = &
\gamma \sum\limits^{}_{j\neq i} \frac{g_{j}}{c^{2}} \int d s_{j}
\frac{\delta\left(\rho_{ij}\right)}
{\gamma_{ji}^{2}} \nonumber\\
& & \left[\left((z^{\mu}_{i} - z^{\mu}_{j})\eta_{\alpha\beta} -
\frac{1}{2}\left(\delta^{\mu}_{\alpha}(z_{i\beta} - z_{j\beta})
+ \delta^{\mu}_{\beta}(z_{i\alpha} - z_{j\alpha})\right)\right)
\left(1 - (\ddot{z}_{j} (z_{i} - z_{j}))\right)\right. \nonumber\\
& & -\left. \left(\dot{z}^{\mu}_{j}\eta_{\alpha\beta} -
\frac{1}{2}(\delta^{\mu}_{\alpha}\dot{z}_{j\beta}
+ \delta^{\mu}_{\beta}\dot{z}_{j\alpha})\right) \gamma_{ji}
\right]
\label{P12}
\end{eqnarray}

It should be emphasized that the action functional (\ref{P6}) is the
most general expression one can obtain from the following four assumptions
\cite{domingo2}:

(1) the action does not depend on the four-vector
accelerations or on higher derivatives,

(2) the interactions propagate at the speed of light in vacuum,

(3) the Poincar\'e invariant parameters that label succesive events along the
world lines of the particles can be identified as their proper times 
in flat spacetime,

(4) all masses are scalars under Poincar\'e transformations.

As it can be seen from (\ref{P6}), significant constraints on the form of the 
action functional result from
these assumptions. The theory depends only on three undetermined constants.
It reduces to the Wheeler-Feynman theory of electrodynamics when two of the
constants ($\alpha$ and $\gamma$) are assumed to be equal to zero.

Based on these results, it is natural to ask how the theory described by the
action (\ref{P6}) relates to physical phenomena that may be observed
experimentally. One possibility is to look at the theory as a modified
theory of electrodynamics and search for effects arising from the new
interaction terms (assuming that $\alpha$ or $\gamma$ are not
exactly equal to zero). Another possibility is to consider the theory as an
incomplete model for the description of strong nuclear interactions (since the
theory incorporates a linearly rising potential, but does not take into account
the internal degrees of freedom of the particles).

The fully relativistic equations of motion (\ref{P8}- \ref{P12}) admit
exact circular solutions for any number of particles. For electrodynamics
such solutions were obtained in \cite{domingo}.

Our purpose now is to obtain, to terms of second order,
the Lagrangian, the Hamiltonian and the equations of motion,
with all dynamical variables depending on a single time $t$
in an inertial reference frame $K$. 
To this end, it is
convenient to rewrite the action functional (\ref{P6}) in the following form:

\begin{equation}
S = 
- \sum\limits^{}_{i} m_{i}c^{2} \int dt_{i}
\left(1 - \frac{v_{i}^{2}}{c^{2}}\right)^{\frac{1}{2}}
- \frac{c}{2} \sum\limits^{}_{i}  \sum\limits^{}_{j\neq i} g_{i} g_{j}
\int \int d t_{i} d t_{j}
\delta\left(c^{2}(t_{i} - t_{j})^{2} - (\vec{r}_{i} - \vec{r}_{j})^{2}\right)
F_{ij}
\label{P14}
\end{equation}

\noindent where,

\begin{eqnarray}
F_{ij} & = & \alpha \left(c (t_{i} - t_{j})
- \frac{(\vec{v}_{j} (\vec{r}_{i} - \vec{r}_{j}))}{c}\right)
\left(c (t_{j} - t_{i})
- \frac{(\vec{v}_{i} (\vec{r}_{j} - \vec{r}_{i}))}{c}\right)
\nonumber\\
& & + \beta \left(1 - \frac{(\vec{v}_{i} \vec{v}_{j})}{c^{2}}\right)
+ \gamma 
\left(1 - \frac{v_{i}^{2}}{c^{2}}\right)^{\frac{1}{2}}
\left(1 - \frac{v_{j}^{2}}{c^{2}}\right)^{\frac{1}{2}}
\label{P15}
\end{eqnarray}

From (\ref{P14}, \ref{P15}) we can obtain the action for an individual
particle $i$,
assuming a fixed motion of the other particles, as follows:

\begin{equation}
S_{i} = 
\int dt \left[
-  m_{i} c^{2} \left(1 - \frac{v_{i}^{2}}{c^{2}}\right)^{\frac{1}{2}}
- c g_{i} \sum\limits^{}_{j\neq i} g_{j}
\int d t_{j}
\delta\left(c^{2}(t - t_{j})^{2} - (\vec{r}_{i} - \vec{r}_{j})^{2}\right)
F_{ij}\right]
\label{P16}
\end{equation}

\noindent where in (\ref{P16}) $F_{ij}$ depends on $t \equiv t_{i}$ and
on $t_{j}$ ($j \neq i$).

The Dirac delta function can be expressed as follows \cite{barut}:

\begin{equation}
\delta\left(c^{2}(t - t_{j})^{2} - (\vec{r}_{i} - \vec{r}_{j})^{2}\right)
= \frac{1}{2 c} \left(
\frac{\delta(t_{j} - t^{(i, -)}_{j})}
{\left(R^{ret}_{ij}
- \frac{\left(\vec{R}^{ret}_{ij} \vec{v}^{(-)}_{j}\right)}{c}\right)}
+
\frac{\delta(t_{j} - t^{(i, +)}_{j})}
{\left(R^{adv}_{ij}
+ \frac{\left(\vec{R}^{adv}_{ij} \vec{v}^{(+)}_{j}\right)}{c}\right)}
\right)
\label{P17}
\end{equation}

In (\ref{P17}), $t^{(i, s)}_{j}$ ($s = -, +$) are the two roots
of the equation:

\begin{equation}
c^{2}(t - t_{j})^{2} - (\vec{r}_{i}(t) - \vec{r}_{j}(t_{j}))^{2} = 0
\label{P18}
\end{equation}

\noindent and,

\begin{equation}
R^{ret}_{ij} = c \left(t - t^{(i,-)}_{j}\right)
\label{P19}
\end{equation}

\begin{equation}
R^{adv}_{ij} = c \left(t^{(i,+)}_{j} - t\right)
\label{P20}
\end{equation}

\begin{equation}
\vec{R}^{ret}_{ij} = \vec{r}_{i} - \vec{r}^{(-)}_{j} 
\label{P21}
\end{equation}

\begin{equation}
\vec{R}^{adv}_{ij} = \vec{r}_{i} - \vec{r}^{(+)}_{j}
\label{P22}
\end{equation}

$t - t^{(i,-)}_{j}$ is the time it takes for a signal to travel forward in
time at the speed of light from particle $j$ to particle $i$ in $K$.

$t^{(i,+)}_{j} - t$ is the time it takes for a signal to travel backward in
time at the speed of light from particle $j$ to particle $i$ in $K$.

The action functional (\ref{P16}) for particle $i$ (assuming a fixed
motion of the other particles) can therefore be rewritten as:

\begin{equation}
S_{i} = 
\int dt \left[
-  m_{i} c^{2} \left(1 - \frac{v_{i}^{2}}{c^{2}}\right)^{\frac{1}{2}}
- \frac{g_{i}}{2} \sum\limits^{}_{j\neq i} g_{j}
\left(
\frac{F^{(-)}_{ij}}
{\left(R^{ret}_{ij}
- \frac{\left(\vec{R}^{ret}_{ij} \vec{v}^{(-)}_{j}\right)}{c}\right)}
+ \frac{F^{(+)}_{ij}}
{\left(R^{adv}_{ij}
+ \frac{\left(\vec{R}^{adv}_{ij} \vec{v}^{(+)}_{j}\right)}{c}\right)}
\right)
\right]
\label{P23}
\end{equation}

\noindent where,

\begin{eqnarray}
F^{(-)}_{ij} & = &  - \alpha
\left(R^{ret}_{ij}
- \frac{\left(\vec{R}^{ret}_{ij} \vec{v}^{(-)}_{j}\right)}{c}\right)
\left(R^{ret}_{ij}
- \frac{\left(\vec{R}^{ret}_{ij} \vec{v}_{i}\right)}{c}\right)
\nonumber\\
& & + \beta \left(1 - \frac{(\vec{v}_{i} \vec{v}^{(-)}_{j})}{c^{2}}\right)
+ \gamma 
\left(1 - \frac{v_{i}^{2}}{c^{2}}\right)^{\frac{1}{2}}
\left(1 - \frac{v^{(-) 2}_{j}}{c^{2}}\right)^{\frac{1}{2}}
\label{P24}
\end{eqnarray}

\begin{eqnarray}
F^{(+)}_{ij} & = &  - \alpha
\left(R^{adv}_{ij}
+ \frac{\left(\vec{R}^{adv}_{ij} \vec{v}^{(+)}_{j}\right)}{c}\right)
\left(R^{adv}_{ij}
+ \frac{\left(\vec{R}^{adv}_{ij} \vec{v}_{i}\right)}{c}\right)
\nonumber\\
& & + \beta \left(1 - \frac{(\vec{v}_{i} \vec{v}^{(+)}_{j})}{c^{2}}\right)
+ \gamma 
\left(1 - \frac{v_{i}^{2}}{c^{2}}\right)^{\frac{1}{2}}
\left(1 - \frac{v^{(+) 2}_{j}}{c^{2}}\right)^{\frac{1}{2}}
\label{P25}
\end{eqnarray}

To terms of second order we can write:

\begin{eqnarray}
& & -  m_{i} c^{2} \left(1 - \frac{v_{i}^{2}}{c^{2}}\right)^{\frac{1}{2}}
- \frac{g_{i}}{2} \sum\limits^{}_{j\neq i} g_{j}
\left(
\frac{F^{(-)}_{ij}}
{\left(R^{ret}_{ij}
- \frac{\left(\vec{R}^{ret}_{ij} \vec{v}^{(-)}_{j}\right)}{c}\right)}
+ \frac{F^{(+)}_{ij}}
{\left(R^{adv}_{ij}
+ \frac{\left(\vec{R}^{adv}_{ij} \vec{v}^{(+)}_{j}\right)}{c}\right)}
\right)
\approx
\nonumber\\
& & 
 - m_{i} c^{2} + \frac{m_{i} v_{i}^{2}}{2}
+ \frac{m_{i} v_{i}^{4}}{8 c^{2}}
- g_{i} \sum\limits^{}_{j\neq i} g_{j} \left[
- \alpha r_{ij} \left(
1 - \frac{(\vec{v}_{i} \vec{v}_{j})}{c^{2}}
+ \frac{(\vec{n}_{ij} \vec{v}_{j})^{2}}{2 c^{2}}
+ \frac{v_{j}^{2}}{2 c^{2}}
- \frac{(\vec{r}_{ij} \vec{a}_{j})}{2 c^{2}}
\right) \right.
\nonumber\\
& & + \frac{\beta}{r_{ij}} \left(
1 - \frac{(\vec{v}_{i} \vec{v}_{j})}{c^{2}}
- \frac{(\vec{n}_{ij} \vec{v}_{j})^{2}}{2 c^{2}}
+ \frac{v_{j}^{2}}{2 c^{2}}
- \frac{(\vec{r}_{ij} \vec{a}_{j})}{2 c^{2}}
\right)
\nonumber\\
& & \left. + \frac{\gamma}{r_{ij}} \left(1 - \frac{v_{i}^{2}}{2 c^{2}}
- \frac{(\vec{n}_{ij} \vec{v}_{j})^{2}}{2 c^{2}}
- \frac{(\vec{r}_{ij} \vec{a}_{j})}{2 c^{2}} \right) \right]
\label{P27}
\end{eqnarray}

We find the
approximate action for particle $i$ to be
(integration by parts allows to
obtain a formula without explicit dependence on the accelerations):

\begin{equation}
S_{i} \approx \int d t L_{i}
\label{P28}
\end{equation}

\noindent where,

\begin{eqnarray}
L_{i} & = & - m_{i} c^{2}  + 
\frac{m_{i} v_{i}^{2}}{2} +  \frac{m_{i} v_{i}^{4}}{8 c^{2}}
+ g_{i} \sum\limits^{}_{j \neq i}
g_{j} \left( \alpha r_{ij} - \frac{(\beta + \gamma)}{r_{ij}} \right)
\nonumber\\
& & + \frac{g_{i}}{2 c^{2}} \sum\limits^{}_{j \neq i} g_{j}
\left[\left(- \alpha r_{ij} + \frac{(\beta - \gamma)}{r_{ij}}\right)
(\vec{v}_{i} \vec{v}_{j}) 
+ \left(\alpha r_{ij} + \frac{(\beta + \gamma)}{r_{ij}}\right)
(\vec{n}_{ij} \vec{v}_{i})(\vec{n}_{ij} \vec{v}_{j})
\right. \nonumber\\
& & \left. + \frac{\gamma}{r_{ij}}\left(v_{i}^{2} + v_{j}^{2}\right)
\right]
\label{P29}
\end{eqnarray}

From (\ref{P29}) we can find the Lagrangian of the whole system of
$N$ interacting particles. It is given by the expression:

\begin{eqnarray}
L & = & - c^{2}\sum\limits^{}_{i} m_{i} +
\sum\limits^{}_{i}
\frac{m_{i} v_{i}^{2}}{2}
+ \sum\limits^{}_{i}
\frac{m_{i} v_{i}^{4}}{8 c^{2}}
+ \frac{1}{2} \sum\limits^{}_{i} \sum\limits^{}_{j \neq i} g_{i} g_{j}
\left(\alpha r_{ij} - \frac{(\beta + \gamma)}{r_{ij}}\right) \nonumber\\
& & + \frac{1}{4 c^{2}}
\sum\limits^{}_{i} \sum\limits^{}_{j \neq i} g_{i} g_{j}
\left[\left(- \alpha r_{ij} + \frac{(\beta - \gamma)}{r_{ij}}\right)
(\vec{v}_{i} \vec{v}_{j}) 
+ \left(\alpha r_{ij} + \frac{(\beta + \gamma)}{r_{ij}}\right)
(\vec{n}_{ij} \vec{v}_{i})(\vec{n}_{ij} \vec{v}_{j})
+ \frac{\gamma}{r_{ij}}\left(v_{i}^{2} + v_{j}^{2}\right)
\right] \nonumber\\
\label{P30}
\end{eqnarray}

The Lagrangian (\ref{P30}) reduces to the Darwin Lagrangian (\ref{P1}) if
$\alpha = \gamma = 0$ and $\beta = 1$.

From the Lagrangian (\ref{P30}) one can obtain the equations of motion to
terms of second order:

\begin{eqnarray}
& & \left(m_{i} + \frac{\gamma g_{i}}{c^{2}}\sum\limits^{}_{j \neq i}
\frac{g_{j}}{r_{ij}}\right)
\left(\left(1 + \frac{v_{i}^{2}}{2 c^{2}}\right)\vec{a}_{i} +
\frac{(\vec{v}_{i}\vec{a}_{i})}{c^{2}}\vec{v}_{i}\right) = \nonumber\\
&  & g_{i} \sum\limits^{}_{j \neq i} \frac{g_{j}}{r_{ij}}
\left[\vec{r}_{ij}\left[
\left(\alpha + \frac{(\beta + \gamma)}{r_{ij}^{2}}\right)
\left(1 - \frac{(\vec{r}_{ij}\vec{a}_{j})}{2 c^{2}}\right)
- \frac{\gamma}{r_{ij}^{2}}\frac{v_{i}^{2}}{2 c^{2}}\right. \right.
\nonumber\\
& & \left. +
\left(\alpha + \frac{\beta}{r_{ij}^{2}}\right)
\left(\frac{v_{j}^{2}}{2 c^{2}}
- \frac{(\vec{v}_{i} \vec{v}_{j})}{c^{2}}\right)
-\left(\alpha + \frac{3 (\beta + \gamma)}{r_{ij}^{2}}\right)
\frac{(\vec{n}_{ij}\vec{v}_{j})^{2}}{2 c^{2}}
\right] \nonumber\\
& & + \left(\alpha + \frac{\beta}{r_{ij}^{2}}\right)
\frac{\vec{v}_{j}(\vec{r}_{ij}\vec{v}_{i})}{c^{2}}  
+ \frac{\gamma}{r_{ij}^{2}}
\left(\frac{\vec{v}_{i}(\vec{r}_{ij}\vec{v}_{i} -\vec{v}_{j})}{c^{2}}
+ \frac{\vec{v}_{j}(\vec{r}_{ij}\vec{v}_{j})}{c^{2}}\right) \nonumber\\
& & \left. + \left(\alpha + \frac{(\gamma - \beta)}{r_{ij}^{2}}\right)
\frac{\vec{a}_{j} r_{ij}^{2}}{2 c^{2}}
\right]
\label{P31}
\end{eqnarray}

The approximate equations (\ref{P31}) can also be derived directly from
the exact equations of motion (\ref{P8} - \ref{P12}).

From the Lagrangian (\ref{P30}) we can obtain the total energy ($E$),
the total linear momentum ($\vec{P}$) and the total angular momentum
($\vec{L}$), 
of a system of $N$
interacting relativistic particles, to terms of second order:

\begin{eqnarray}
E & = & c^{2}\sum\limits^{}_{i} m_{i} + \sum\limits^{}_{i} 
\frac{m_{i} v_{i}^{2}}{2}
+ \sum\limits^{}_{i}
\frac{3 m_{i} v_{i}^{4}}{8 c^{2}}
- \frac{1}{2} \sum\limits^{}_{i} \sum\limits^{}_{j \neq i} g_{i} g_{j}
\left(\alpha r_{ij} - \frac{(\beta + \gamma)}{r_{ij}}\right) \nonumber\\
& & +  \frac{1}{4 c^{2}}
\sum\limits^{}_{i} \sum\limits^{}_{j \neq i} g_{i} g_{j}
\left[\left(- \alpha r_{ij} + \frac{(\beta - \gamma)}{r_{ij}}\right)
(\vec{v}_{i} \vec{v}_{j}) 
 + \left(\alpha r_{ij} + \frac{(\beta + \gamma)}{r_{ij}}\right)
(\vec{n}_{ij} \vec{v}_{i})(\vec{n}_{ij} \vec{v}_{j})
+ \frac{\gamma}{r_{ij}}\left(v_{i}^{2} + v_{j}^{2}\right)
\right] \nonumber\\
\label{P32}
\end{eqnarray}

\begin{eqnarray}
\vec{P} & = & \sum\limits^{}_{i} \left(m_{i} 
\left(1 + \frac{v_{i}^{2}}{2 c^{2}}\right)
+ \frac{\gamma g_{i}}{c^{2}}
\sum\limits^{}_{j \neq i} \frac{g_{j}}{r_{ij}}\right) \vec{v}_{i} \nonumber\\
& & + \frac{1}{2 c^{2}} \sum\limits^{}_{i} \sum\limits^{}_{j \neq i}
g_{i} g_{j} \left[
\left( -\alpha r_{ij} + \frac{(\beta - \gamma)}{r_{ij}}\right)\vec{v}_{j}
+  \left(\alpha r_{ij} + \frac{(\beta + \gamma)}{r_{ij}}\right)
(\vec{n}_{ij} \vec{v}_{j}) \vec{n}_{ij}
\right]
\label{P33}
\end{eqnarray}

\begin{eqnarray}
\vec{L} & = &
\sum\limits^{}_{i} \left(m_{i}\left(1 + \frac{v_{i}^{2}}{2 c^{2}}\right)
+ \frac{\gamma g_{i}}{c^{2}}\sum\limits^{}_{j \neq i} \frac{g_{j}}{r_{ij}}
\right) (\vec{r}_{i} \times \vec{v}_{i})\nonumber\\
& & + \frac{1}{2 c^{2}} \sum\limits^{}_{i} \sum\limits^{}_{j \neq i}
g_{i} g_{j} \left[
\left(- \alpha r_{ij} + \frac{(\beta - \gamma)}{r_{ij}}\right)
(\vec{r}_{i} \times \vec{v}_{j})
- \left(\alpha r_{ij} + \frac{(\beta + \gamma)}{r_{ij}}\right)
\frac{(\vec{n}_{ij} \vec{v}_{j})}{r_{ij}} (\vec{r}_{i} \times \vec{r}_{j})
\right]
\nonumber\\
\label{P34}
\end{eqnarray}

In order to develop the hamiltonian formulation of the theory,
we find the conjugate momenta:

\begin{eqnarray}
\vec{p}_{i}  = \frac{\partial L}{\partial \vec{v}_{i}} & = &
m_{i} \left(1 + \frac{v_{i}^{2}}{2 c^{2}}
+ \frac{\gamma g_{i}}{m_{i} c^{2}} \sum\limits^{}_{j \neq i}
\frac{g_{j}}{r_{ij}}\right) \vec{v}_{i} \nonumber\\
& & + \frac{g_{i}}{2 c^{2}} \sum\limits^{}_{j \neq i} g_{j}
\left[
\left(
- \alpha r_{ij} + \frac{(\beta - \gamma)}{r_{ij}}
\right) \vec{v}_{j} +
\left(
\alpha r_{ij} + \frac{(\beta + \gamma)}{r_{ij}}
\right)(\vec{n}_{ij} \vec{v}_{j}) \vec{n}_{ij}
\right]
\label{P35a}
\end{eqnarray}

The Hamiltonian (to terms of second order) for $N$ interacting particles
is given by the formula:

\begin{eqnarray}
H & = & c^{2}\sum\limits^{}_{i} m_{i}
+ \sum\limits^{}_{i} \frac{p_{i}^{2}}{2 m_{i}}
- \sum\limits^{}_{i} \frac{p_{i}^{4}}{8 m_{i}^{3} c^{2}}
- \frac{1}{2} \sum\limits^{}_{i} \sum\limits^{}_{j \neq i} g_{i} g_{j}
\left(\alpha r_{ij} - \frac{(\beta + \gamma)}{r_{ij}}\right) \nonumber\\
& & - \frac{1}{4 c^{2}} \sum\limits^{}_{i} \sum\limits^{}_{j \neq i}
\frac{g_{i} g_{j}}{m_{i} m_{j}} \left[
\left(- \alpha r_{ij} + \frac{(\beta - \gamma)}{r_{ij}}\right)
(\vec{p}_{i} \vec{p}_{j}) 
+ \left(\alpha r_{ij} + \frac{(\beta + \gamma)}{r_{ij}}\right)
(\vec{n}_{ij} \vec{p}_{i})(\vec{n}_{ij} \vec{p}_{j})
\right] \nonumber\\
& & - \frac{\gamma}{4 c^{2}}
\sum\limits^{}_{i} \sum\limits^{}_{j \neq i}
\frac{g_{i} g_{j}}{r_{ij}} \left(\frac{p_{i}^{2}}{m_{i}^{2}}
+  \frac{p_{j}^{2}}{m_{j}^{2}} \right) \nonumber\\
\label{P35}
\end{eqnarray}

For $N = 2$ in the center of momentum frame (c.m.f.)
($\vec{p}_{1} = - \vec{p}_{2} = \vec{p}$) the Hamiltonian (\ref{P35}) takes
the simple form:

\begin{eqnarray}
H & = & c^{2} \left(m_{1} + m_{2}\right)
+ \frac{p^{2}}{2} \left(\frac{1}{m_{1}} + \frac{1}{m_{2}}\right)
- g_{1} g_{2} \left(\alpha r - \frac{(\beta + \gamma)}{r}\right)
\nonumber\\
& & - \frac{p^{4}}{8 c^{2}} \left(\frac{1}{m_{1}^{3}} + \frac{1}{m_{2}^{3}}\right)
- \frac{g_{1} g_{2}}{2 c^{2}}
\left(\frac{1}{m_{1}^{2}} + \frac{1}{m_{2}^{2}}\right) \frac{\gamma p^{2}}{r}
\nonumber\\
& & + \frac{g_{1} g_{2}}{2 c^{2} m_{1} m_{2}}
\left(\left(- \alpha r + \frac{(\beta - \gamma)}{r}\right) p^{2}
+ \left(\alpha r + \frac{(\beta + \gamma)}{r}\right)(\vec{n} \vec{p})^{2}\right)
\label{P36}
\end{eqnarray}

Direct predictions for the energy spectrum can be made from (\ref{P36}).

Comparing our expression for the hamiltonian (\ref{P36}) with the general expression 
presented in \cite{childers}, we can determine the Lorentz structure of the
interactions. 

The general formula for the semirelativistic two-particle hamiltonian
presented in \cite{childers} can be written in the c.m.f. as follows:

\begin{equation}
H = c^{2} \left(m_{1} + m_{2}\right)
+ \frac{p^{2}}{2} \left(\frac{1}{m_{1}} + \frac{1}{m_{2}}\right)
- \frac{p^{4}}{8 c^{2}} \left(\frac{1}{m_{1}^{3}} + \frac{1}{m_{2}^{3}}\right) + W
\label{R1}
\end{equation}

\noindent where $W$ is the sum of vector and scalar interactions:

\begin{equation}
W = W_{V} + W_{S}
\label{R2}
\end{equation}

Given a potential $V$ for vector exchange, $W_{V}$ in the c.m.f. is given by the formula:

\begin{equation}
W_{V} = V(r) + \frac{1}{2 m_{1}m_{2}c^{2}} \left(V(r) p^{2} - r V'(r) (\vec{n}\vec{p})^{2}\right)
\label{R3}
\end{equation}

Given a potential $S$ for scalar exchange, $W_{S}$ in the c.m.f. can be written as follows:

\begin{equation}
W_{S} = S(r) - S(r) \left(\frac{1}{m_{1}^{2}} + \frac{1}{m_{2}^{2}}\right) \frac{p^{2}}{2 c^{2}} 
- \frac{1}{2 m_{1}m_{2}c^{2}} \left(S(r) p^{2} + r S'(r) (\vec{n}\vec{p})^{2}\right)
\label{R4}
\end{equation}

Comparing (\ref{P36}) with (\ref{R1} - \ref{R4}) we conclude that,
for the theory presented in this Letter, the potentials corresponding to vector and scalar interactions are 
given by the following expressions:

\begin{equation}
V =  g_{1} g_{2} \left(- \alpha r + \frac{\beta}{r}\right)
\label{R5}
\end{equation}

\begin{equation}
S = \frac{g_{1} g_{2} \gamma}{r}
\label{R6}
\end{equation}

The interaction responsible for the variable masses is scalar in nature.
The lineraly rising potential is purely a vector potential, and so is the Coulomb-like potential.
All this can also be seen directly from the fully relativistic action (\ref{P6}).

For $N = 2$, the approximate equations of motion(\ref{P31}) admit circular
orbits solutions
with the angular frequency $\omega$ and the distance between the
particles $r = r_{1} + r_{2}$ obeying the generalized Kepler relation:

\begin{equation}
\omega^{2} = - \frac{g_{1} g_{2}}{\mu r^{3}}
\left(\alpha r^{2} + \beta + \gamma \right)
\left[1 - \frac{g_{1} g_{2}}{\mu c^{2} r}
\left(3\alpha \nu r^{2} + 
\beta\left(\nu - \frac{1}{2}\right) + \gamma \nu \right)
\right]
\label{P37}
\end{equation}

\noindent where,

\begin{equation}
\mu = \frac{m_{1} m_{2}}{m_{1} + m_{2}}
\label{P38}
\end{equation}

\begin{equation}
\nu = \frac{m_{1} m_{2}}{(m_{1} + m_{2})^{2}}
\label{P39}
\end{equation}

For two-body circular orbits the total energy of the system in the c.m.f.,
to terms of second order, is given by the formula:

\begin{equation}
E = c^{2} (m_{1} + m_{2}) +
\frac{g_{1} g_{2}}{2 r} \left[
- 3\alpha r^{2} + \beta + \gamma + \frac{g_{1} g_{2}}{4 c^{2} \mu r}
\left[
(\alpha r^{2} + \beta)^{2} (1 - \nu) - 2 (\alpha r^{2} + \beta) \gamma \nu
-\gamma^{2} (1 + \nu) \right]\right]
\label{P40}
\end{equation}

The total angular momentum in the c.m.f, to terms of second order, is:

\begin{equation}
L = \left|
g_{1} g_{2} (\alpha r^{2} + \beta +\gamma) \mu r \right|^{\frac{1}{2}}
\left[
1 + \frac{g_{1} g_{2}}{4 c^{2} \mu r}\left[4 \alpha \nu r^{2}
- \beta + 2 \gamma \right]
\right]
\label{P41}
\end{equation}

One can use the procedure presented in this paper to obtain results at
higher orders in the series expansions.

At the quantum level one can describe a system of two spinning
particles interacting electromagnetically, including second order terms,
by the Breit equation \cite{breit}- \cite{2bDirac}.
An extended Breit equation may be devised in agreement with the
Lagrangian (\ref{P30})
to account for the dynamical
contributions to the masses and the effect of a linearly rising
potential.

It may also be possible to obtain a fully relativistic two-body Dirac equation
from the action (\ref{P6}),
by extending the results of Crater and Van Alstine \cite{2bDirac} to include
the additional interactions explored in this Letter.

\end{document}